\documentclass[lettersize,journal]{IEEEtran}
\usepackage{subfigure}
\usepackage{amsmath,amsfonts}
\usepackage{soul}
\usepackage{array}
\usepackage[caption=false,font=normalsize,labelfont=sf,textfont=sf]{subfig}
\usepackage{textcomp}
\usepackage{caption}
\usepackage{stfloats}
\usepackage{url}
\usepackage{subcaption}
\usepackage{verbatim}
\usepackage{graphicx}
\usepackage{float}
\usepackage{algorithm}
\usepackage{dirtytalk}
\usepackage{enumitem}

\usepackage{multicol}
\usepackage{multirow}
\usepackage{booktabs}

\usepackage{nicefrac}
\usepackage{algpseudocode}
\def\BibTeX{{\rm B\kern-.05em{\sc i\kern-.025em b}\kern-.08em
    T\kern-.1667em\lower.7ex\hbox{E}\kern-.125emX}}
\usepackage{balance}

\IEEEpubid{\begin{minipage}{\textwidth}\centering \scriptsize © 2026 IEEE. All rights reserved, including rights for text and data mining and training of artificial intelligence and similar technologies. Personal use is permitted, but republication/redistribution requires IEEE permission. See \url{https://www.ieee.org/publications/rights/index.html} for more information.\end{minipage}}

\begin{document}
\title{Beyond the Utterance: An Empirical Study of Very Long Context Speech Recognition}
\author{Robert Flynn, Anton Ragni
\thanks{
Both authors are with the Department of Computer Science, The University Of Sheffield, United Kingdom (e-mail: \{rjflynn2, a.ragni\}@sheffield.ac.uk).

This work was supported by the CDT in Speech and Language Technologies (SLT) and their Applications funded by UKRI [grant number
EP/S023062/1].}} 

\markboth{transactions on audio speech and language processing}%
{How to Use the IEEEtran \LaTeX \ Templates}

\maketitle
    
    \begin{abstract}
    Automatic speech recognition (ASR) models are normally trained to operate over single utterances, with a short duration of less than 30 seconds. This choice has been made in part due to computational constraints, but also reflects a common, but often inaccurate, modeling assumption that treats utterances as independent and identically distributed samples. When long-format audio recordings are available, to work with such systems, these recordings must first be segmented into short utterances and processed independently. In this work, we show that due to recent algorithmic and hardware advances, this is no longer necessary, and current attention-based approaches can be used to train ASR systems that operate on sequences of over an hour in length. Therefore, to gain a better understanding of the relationship between the training/evaluation sequence length and performance, we train ASR models on large-scale data using 10 different sequence lengths from 10 seconds up to 1 hour. The results show a benefit from using up to 21.8 minutes of context, with up to a 14.2\% relative improvement from a short context baseline in our primary experiments. Through modifying various architectural components, we find that the method of encoding positional information and the model's width/depth are important factors when working with long sequences. Finally, a series of evaluations using synthetic data are constructed to help analyse the model's use of context. From these results, it is clear that both linguistic and acoustic aspects of the distant context are being used by the model.

    \end{abstract}
    
    
    
    \section{Introduction}
    \IEEEpubidadjcol
    
    \IEEEPARstart{M}{any} important use-cases for Automatic Speech Recognition (ASR) involve long-format audio, such as meeting and lectures. \textit{Ideally}, models would be able to process audio sequences of arbitrary length in an end-to-end fashion, and adapt based on all of the provided data. However, current ASR models are generally trained to operate on short sequence lengths of around 5-30s \cite{hori2021advanced, lu2021input, whisperradford2023robust, chiu2019comparison}. This design choice is made in part due to computational constraints, as the self-attention operator \cite{vaswani2017attention} that acts as the backbone for state-of-the-art ASR systems \cite{whisperradford2023robust, gulati2020conformer}, has quadratic computational complexity with respect to the sequence lengths. However, with modern hardware and algorithmic improvements \cite{dao2022flashattention, liu2023ring}, the quadratic complexity of self-attention becomes continually less limiting. For example, recent work \cite{howmuchcontextflynn} demonstrates that it is possible to train Conformer-based ASR models on sequence lengths of 1 hour. 
    
    However, it is not clear whether the current self-attention based architectures are able to fully utilise the information in longer acoustic sequences, with \cite{howmuchcontextflynn} reporting a plateau in word error rate reduction for sequences longer than 21.8 minutes. Similar findings have been reported for language modelling on WikiText \cite{press-etal-2021-shortformer} where improvements in perplexity plateau after 1024 tokens, and long-context toy tasks such as Path-256 \cite{tay2020long} where training completely fails. As such, this work looks to provide a comprehensive study on long-context speech recognition that can aid researches and practitioners looking to research or build ASR models for long-format applications. 
    A list of our research questions alongside a summary of our key contributions is given as follows: 
    \begin{itemize}[label={}, leftmargin=*]
        \item \textbf{How much context do ASR systems benefit from?} We show ($\S$\ref{sec:howmuchcontext}) that the ASR systems investigated in this work benefit from up to 21.8 minutes of context. 
        
        \item \textbf{What architectural or training modifications improve or worsen the ASR system's use of long-range context?} We show ($\S$\ref{sec:impactofpos}-$\S$\ref{sec:impactofmodelsize}) that altering factors such as the number of layers, model width and the positional encoding scheme can impact the model's ability to benefit from longer contexts. Increasing the total parameter count or number of training epochs ($\S$\ref{sec:impactofepoechs}) improves overall model performance but does not improve the model's use of long-range context relative to a short context baseline.
        
        \item \textbf{What are the settings where long contexts are most beneficial?} We evaluate the models on a range of different domains, using 4 different evaluation datasets. We find ($\S$\ref{sec:howmuchcontext}) that training/evaluating with a longer context improves model performance most when the domain shift between training and testing is the largest. For in-domain data there was no meaningful benefit from using context sizes beyond 20–82 seconds. We also found that longer context models are more robust to the addition of background noise.

        \item \textbf{What aspects of the context is the ASR system benefitting from?} We construct several synthetic evaluations that show ($\S$\ref{sec:howusecontext}) that the models are using both linguistic and acoustic components of the long-range context to inform the predictions. Although, we find ($\S$\ref{sec:incontexteval}) evidence that the model's linguistic understanding of the context is fairly simplistic.
    \end{itemize}

All of the code used for training/evaluation, along with the results and checkpoints for all models, is made available here\footnote{\url{https://github.com/robflynnyh/long-context-asr}}.

\IEEEpubidadjcol

\section{Prior Work}
\subsection{Long-Context Acoustic Models}
The majority of the literature on long-context acoustic models (AMs) still uses a relatively short context.  
For example, \cite{hori2020transformer, hori2021advanced} look at extending the context window of encoder decoder ASR models by including keys/value from previous utterances. However, the previous context is limited to around 20–25 seconds. Fusing context from previous utterance using an attention pooling mechanism is explored in \cite{cui2023towards}. This investigation is limited to a context of 2 preceeding utterances. 
Other work \cite{carvalho2023memory} uses the external memory from
Neural Turing Machine \cite{graves2014neural} to share context information between frames. However, only context sizes of 1–20 seconds are used during training, with a maximum sequence length of 70 seconds during evaluation. Performance at different context sizes using a FastConformer based architecture \cite{rekesh2023fastconformer} is examined in \cite{koluguri2024longer}. However, the maximum sequence length considered in their training and evaluation setup is 60 seconds. The widely known Whisper work \cite{whisperradford2023robust} investigates ASR over long-form recordings, but the model is always limited to an acoustic context of 30 seconds and 448 previous text tokens. 

Recordings of an hour or more in length are used when evaluating ASR models in \cite{koluguri2023investigating, rekesh2023fastconformer}. However, local/windowed attention with a window size of +/- 10 seconds is employed, which prevents the model from using the very long range context. They also investigate the use of a global token; however, the model is only trained on sequences with a maximum length of 20s causing a potential train/test mismatch. Additionally, it is not clear how much context the global token allows the model to benefit from. The most relevant work to ours is \cite{chen2024train}, which investigates training and evaluating on fairly long recordings. However, only 2 settings are investigated: full-recording (maximum length of around 20 minutes) and utterance-level. Hence, it is not clear whether the model is benefitting from the \textit{entire} 20 minute recording. 

\subsection{Long-Context Language Modelling}
The use of longer inputs has seen more investigation in the field of language modelling, which is likely due to the better availability of long-format text data compared to audio. For Long Short-Term Memory \cite{hochreiter1997longlstm} based language models (LMs) \cite{khandelwal2018sharp} finds that on average the model is able to benefit from 200 previous tokens. However, it is not sensitive to the word order beyond 50 tokens, suggesting only a rough understanding of the distant past. Other work \cite{press-etal-2021-shortformer}, with attention-based networks, trains and evaluates LMs at various sequence lengths on the WikiText dataset \cite{merity2016pointer}. The results showed no benefit from using more than 1024 tokens, as the perplexity plateaued when extending beyond this. Similar results are found in \cite{sun2021long} when testing local \cite{beltagy2020longformer,child2019generating,parmar2018image} and routing \cite{roy2021efficient} transformers, with performance plateauing beyond 1024-2048 tokens. Our investigation was motivated by these works, as there was no consensus whether similar trends existed for AMs.

More recently larger LMs trained on internet-scale data, have shown a benefit from larger context sizes. For example, \cite{olsson2022context} reports plots where perplexity begins to plateu at around 10K tokens, and for programming based data Code Llama \cite{roziere2023code} reports improvements from using up to 100K prior tokens. These results suggest that having a large and diverse training/evaluation dataset is important for investigating long-context models.



\section{Modifications For training with long sequences}
\subsection{Flash Attention}
Self-attention \cite{vaswani2017attention} is one of the few operators that has been shown to be reliably capable of learning many complex dependencies that span arbitrary distances. However, the standard computation of self-attention has quadratic memory (and computational) complexity, which prohibits processing sequences longer than a few thousand frames on most devices. This quadratic memory cost is due to calculating pairwise dot products between two different projections $\mathbf{q}$ and $\mathbf{k}$ of the sequence, resulting in a $\mathbf{q} \times \mathbf{k}$ similarity matrix. Flash Attention \cite{dao2022flashattention} is an efficient graphics processing unit (GPU) kernel for computing self-attention without approximations, that avoids materializing the full $\mathbf{q} \times \mathbf{k}$ matrix, and instead computes it in chunks. This also avoids having to transfer the large $\mathbf{q} \times \mathbf{k}$ matrix between the GPUs high-bandwidth memory and L2 cache memory, resulting in a sizeable speed-up (see figure \ref{fig:conformerfastconformerspeed}).  

\subsection{Architecture}

\begin{figure}
    \centering
    \includegraphics[width=7cm]{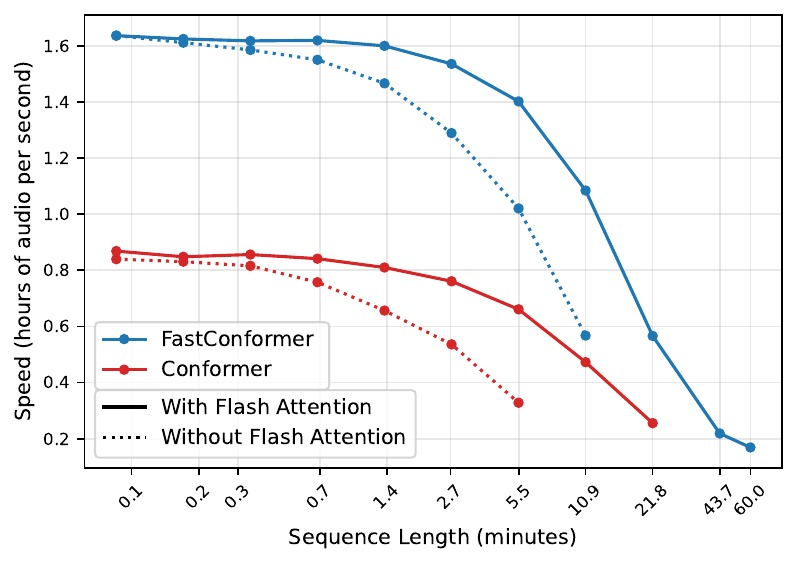}
    \caption{Training throughput (hours of audio processed per second on an H100 GPU) at different sequence lengths.}
    \label{fig:conformerfastconformerspeed}
\end{figure}



For these investigations, we choose to work with encoder-only Conformer \cite{gulati2020conformer} based AMs, that are trained using connectionist temporal classification (CTC) \cite{graves2006connectionist}. CTC models are chosen for their simplicity (and efficiency), as cross-attention mechanisms used in encoder-decoder based approaches may complicate the investigation. The Conformer is a variation of the macaron-style \cite{lu2019understanding} Transformer \cite{vaswani2017attention} that adds a depthwise convolution module after each self-attention branch. A standard implementation of this architecture takes Mel-spectrograms as input and uses a series of 2D convolution subsampling layers prior to the Conformer layers to reduce the initial sequence length by $4\times$. The recent FastConformer \cite{rekesh2023fastconformer} modifies this setup and uses $8\times$ subsampling with 2D depthwise convolutions that have a feature dimension smaller than that of the rest of the model. Results demonstrated that this modification provides a favourable accuracy-efficiency trade-off, and we find that this roughly halves the time per-epoch and dramatically decreases total memory usage. This enables the training of acoustic models with sequence lengths of up to 70 minutes, on 1 80 GB A/H100 GPU when using the Flash Attention algorithm \cite{dao2022flashattention} to compute attention. A comparison displaying the speed difference during training between these two subsampling approaches at various sequence lengths is shown in figure \ref{fig:conformerfastconformerspeed}. As shown, the FastConformer with Flash Attention is the only configuration that is able to fit 1 hour long sequences into memory during training.

\subsection{Sequence Length Warmup}


We found that training from scratch with a sequence length longer than 20–30 seconds resulted in instability that prevented the model from converging. To remedy this, a sequence length warm-up is employed. Specifically, we start training with an initial sequence length $s_0$, which is doubled after every $n$ steps, until a maximum sequence length $s_{\mathrm{max}}$ is reached. The sequence length used at a given step/recording $s_r$ is therefore given as $s_r = \min(s_0 + s_0 \cdot 2 ^{\lfloor \nicefrac{r}{n} \rfloor}, s_{\mathrm{max}}   )$, where $r$ denotes the current recording index (the number of full recordings trained on). 

\subsection{Positional Encoding}
The self-attention operator that is used in Transformer \cite{vaswani2017attention} based models is invariant to permutations of the sequence. These models therefore generally rely on some additional method of encoding the absolute or relative position of tokens/frames. For longer sequences this becomes particularly important as the majority of very distant frames are likely irrelevant. As such, we investigate a range of positional encoding methods in this work, we focus on methods which are currently compatible with Flash Attention \cite{dao2022flashattention}. The methods that we consider are described as follows: 

\subsubsection{No Positional Encodings (NoPos)} 
no explicit method is used, and the model relies on the convolution modules used in Conformer models to encode positional features.
\subsubsection{Sinusoidal Positional Encodings}
were originally introduced in \cite{vaswani2017attention}, this method uses sine/cosine waves to encode the absolute position of tokens/frames. These are typically added to the input at the start of the model. We use the method from \cite{li2021learnable} which functions as follows:
{\small
\begin{equation}
    \mathbf{x}^{'}_i = \mathbf{x}_i + \frac{\mathrm{Concat}(\sin(n_i \mathbf{w}_r), \cos(n_i \mathbf{w}_r))}{\sqrt{d}} 
\end{equation}
}

\noindent where $\mathbf{x}_i \in \mathbb{R}^d$ is a given token/frame at sequence position $n_i$ (where $n_i$ is a scalar integer). $\mathbf{w}_r \in \mathbb{R}^{d/2}$ is a learnable vector, and $d$ is the size of the model's feature dimension.

\subsubsection{Rotary Positional Encodings \cite{su2021roformer}} are applied to the queries and keys prior to the self-attention computation at each layer. This method uses a rotation matrix to encode the absolute and relative position of each token/frame. Specifically, the query $\mathbf{q}$ and key $\mathbf{k}$ vectors are transformed by the block diagonal matrix $\mathbf{R}^{d}_{\Theta, n_i}$, which is shown as follows:

{\small 
\begin{equation}
(\mathbf{R}^{d}_{\Theta, n_i})_{j} = 
\begin{pmatrix} 
 \cos{n_i \theta_j} & -\sin{n_i \theta_j} \\
 \sin{n_i \theta_j} & \cos{n_i \theta_j} \\
\end{pmatrix}
\end{equation}
}

\noindent where $d$ represents the embeddings dimension and $j = \{1,\hdots, \nicefrac{d}{2}\}$, $n_i$ is the absolute position of token $i$. The function $f_R$ that is used to transform the queries $\mathbf{q}$ and keys $\mathbf{k}$ is therefore defined as: $
    f_R(\mathbf{x}_i) =   (\mathbf{R}^{d}_{\Theta, n_i}) \mathbf{x}_i
$, where $\mathbf{x}_i \in \{\mathbf{q}_i, \mathbf{k}_i\}$. The rotation frequency $\theta_j$ is equal to $\theta^{\nicefrac{-2(j-1)}{d}}$, where the base frequency $\theta$ is a hyperparameter that is typically set to a value of $10,000$ following \cite{su2021roformer, vaswani2017attention}. However, increasing $\theta$ has been shown \cite{roziere2023code} to be beneficial when working with long sequences. This may be because it can result in less bias towards nearby tokens in the attention formulation, an in-depth analysis into this effect is provided in \cite{barbero2024round}. Therefore, we  investigate a range of $\theta$ values (from 10k to 10M) and report on a setting where $\theta = 1.5$M.

\begin{figure*}[t!]
  \centering
  \begin{minipage}{.35\textwidth}
    \centering
    \includegraphics[width=7cm]{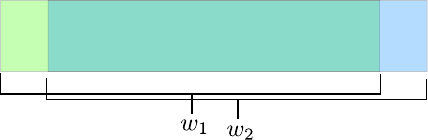}
    \caption{Moving averaged window decoding.}
    \label{fig:movingavgwindow}
  \end{minipage}\hfill
  \begin{minipage}{.35\textwidth}
    \centering
    \includegraphics[width=7cm]{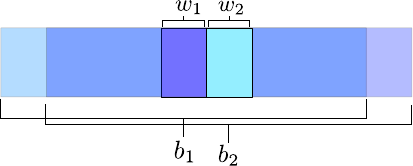}
    \caption{Buffered window decoding.}
  \end{minipage}\hfill
  \begin{minipage}{.21\textwidth}
    \centering
    
    \includegraphics[width=2cm]{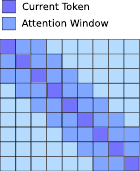}
    \caption{Sliding Window Attention.}
    \label{fig:windowattn}
  \end{minipage}
  \caption*{\textbf{Various approaches for transcribing long-form audio that prevent context fragmentation.}}
\end{figure*}


\section{Modifications For Evaluating with Long Sequences}
\label{sec:contextfrag}

\begin{figure}[ht!]
  \centering
  \begin{subfigure}{}
    \includegraphics[width=5cm]{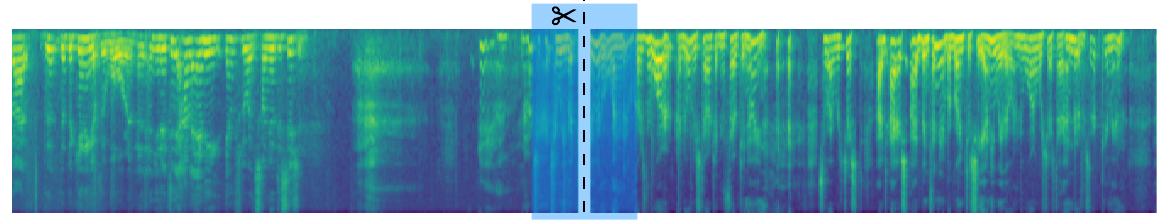}

  \end{subfigure}

  \begin{subfigure}{}
    \includegraphics[width=5cm]{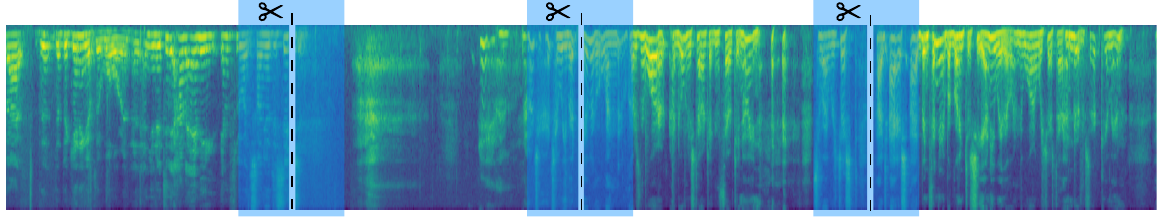}
  \end{subfigure}
  \caption{Depiction of context fragmentation when a recording is segmented. (Top) Sequence length of 10s results in 1 area of fragmentation. (Bottom) Sequence length of 5s results in 3 areas of fragmentation.}
  \label{fig:contextfrag}
\end{figure}

Segmenting long-format audio results in positions where the left or right context is limited, which leads to worse accuracy near the start and end of each sequence. We refer to this as context fragmentation. The use of shorter sequence lengths results in a greater amount of context fragmentation.
In these areas where the context is fragmented, the accuracy of the model's predictions is typically worse. This makes it challenging to compare models that use different amounts of context, as a longer sequence length will also result in less context fragmentation, which is depicted in figure \ref{fig:contextfrag}. Therefore, a naive evaluation where utterances are concatenated to form successively longer sequences can show continually improving performance due to reducing areas of context fragmentation without the model necessarily using any of the distant context. This phenomenon was also explored in \cite{press-etal-2021-shortformer} for the case of autoregressive language modelling.

This is an issue that is often overlooked in many of the previous investigations \cite{chen2024train, koluguri2024longer} into ASR on long-form audio. Which can \textit{potentially} lead to incorrect conclusions about where any improvements from a particular method comes from. Therefore, to fairly compare models that use different sequence lengths, we explore 3 different evaluation schemes that are designed to avoid fragmenting the context. These schemes are depicted in figures \ref{fig:movingavgwindow}-\ref{fig:windowattn}, and are described in the following subsections.

\subsection{Moving Averaged Window Decoding}
\label{sec:movingaveragewindow}
For this method, the spectrogram is processed in windows $\mathcal{W} = \{w_0, \dots, w_K\}$, using a stride of $S$, and the start position of the $k$-th window $w_k$ is given by  $ k \cdot S$. The window width is set to the sequence length of the model. For strides that are shorter than the sequence length this results in multiple predictions per output frame, these are averaged (post-softmax) to obtain the final predictions. We find that a smaller stride is necessary for models that use shorter sequence lengths, and therefore set the stride as a percentage of the sequence length. 

\subsection{Buffered Window Decoding}
\label{sec:bufferedwindow}
For the buffered scheme, we process the spectrogram in segments (buffers) with a width $w_b$ equal to the sequence length of the model. To ensure all predictions are made with both past and future context, only probabilities from a central window $w_w$ of the buffer are used, and the rest are discarded. The buffer is moved forward with a stride equal to $w_w$ so that the start position of the $k$-th buffer is given by $k \cdot w_w - \frac{w_b - w_w}{2}$. The central window width $w_w$ is set to a percentage of the buffer size.

\subsection{Sliding Window Attention}
\label{sec:slidingwindowattention}
For sliding window attention (SWA), the sequence length during evaluation for all models is set to the length of the entire recording. The attention operator for each token is then limited to a local window \cite{beltagy2020longformer,child2019generating,parmar2018image} that is equal to the sequence length used during training. This is performed efficiently using blockwise sparse attention kernels. For devices with less memory, or extremely long recordings, this method can be paired with either of the other decoding methods.

\section{Experimental Configuration}
\subsection{Dastasets}
For these investigations the following datasets are used:
\begin{itemize}[label={}, leftmargin=*]
    \item \textbf{Spotify Podcasts}: For training models, the collection of Spotify podcast provided as part of \cite{spotify-clifton-etal-2020-100000} is used. This includes 105,360 podcasts with a combined length of 58K hours. The average length is around 30 minutes with a maximum length of 300 minutes. The transcripts provided consist of pseudo labels from the Google Speech API.
    \item \textbf{Floras 50}: \cite{chen2024floras} is a corpus of conversational speech collected from YouTube. The data comes pseudo-labelled via Gemini Flash \cite{team2024gemini}. Only the monolingual training split is used in this work. Alignments for this data are attained using CTC models from \cite{pratap2024scaling}. After preparation, the training split contained 30,482 recordings with an average length of 20.4 minutes and a max length of 6.1 hours. This is used as training data for the investigation in \S \ref{sec:floras50}.
    \item \textbf{TED-LIUM}: \cite{hernandez2018ted} is a series of TED talks, only the test set is used, featuring 11 talks, each around 20 minutes long.
    \item \textbf{Earnings-22}: \cite{del2022earnings} consists of earning report meetings and is a challenging dataset with a range of accents. Meetings last around an hour, and for the primary experiments we use the entire dataset as a test set, with 127 hours in total. For the more expensive evaluations detailed in \ref{sec:howusecontext}, the test partition proposed in \cite{gandhi2022esb} is used, which totals 5.6 hours.
    \item \textbf{This American Life Podcast (TAL)}: \cite{mao2020speech} consists of episodes from the radio show TAL. Episodes last 1 hour and include up to 18 speakers and a range of topics are discussed. We use the test split which contains 26 hours of audio.
    \item \textbf{Rev16}: is a test set composed of 16 podcasts form Apple Music, and was originally used in \cite{whisperradford2023robust}. 
\end{itemize}

\subsection{Model Configurations}
All models use an encoder-only Conformer architecture \cite{gulati2020conformer}, with self-conditioned CTC \cite{nozaki2021relaxingselfconditioned} (without intermediate losses). Batch normalisation \cite{ioffe2015batch} is replaced with batch renormalisation \cite{ioffe2017batchrenorm}, to enable non-IID training settings where the sequence length is large, but the batch size is small. For the Conformer's convolution modules, a kernel size of 9 is used. All models use the same 4095 byte-pair-encoding vocabulary, learnt from the training corpus, with an additional token for the blank character used for CTC. For all experiments excluding the subsampling investigation in $\S$\ref{sec:ssfactor}, the Fast Conformer \cite{rekesh2023fastconformer} 8$\times$ subsampling configuration is used.

\subsection{Training Configuration}
\label{sec:trainingconfig}
All models are trained using the Madgrad \cite{defazio2022adaptivitymadgrad} optimizer\footnote{\scriptsize We found this performed better than Adam(W), particularly at smaller batch sizes} and a cosine annealing learning rate schedule with a warm-up. The following context sizes (in seconds) are investigated for the experiments: 10, 20, 41, 82, 164, 328, 655, 1311, 2621 and 3600. For training, the recordings are partitioned into chunks equal to the sequence length and word-level time-steps are used to retrieve the corresponding text. A separate model is trained for each sequence length. All models are trained on the Spotify data except those in \S \ref{sec:floras50}, which use Floras-50.

All experiments use a sequence length warm-up with the hyperparameters $s_0 = 5.12$s $n = 5000$
To fairly compare models trained with different sequence lengths, we aim to keep the total number of training steps roughly the same. In our primary experiments (all results excluding those in figures \ref{fig:wer_per_area_e22_epoch1}, \ref{fig:wer_per_epoch_per_seq} and \ref{fig:wer_per_area_epoch6} and table \ref{tab:floras50}) we accomplish this by varying $s_{max}$ based on the context length and halving the batch size when the sequence length is doubled to maintain a max batch duration of 1 hour of audio.
However, this results in different batching schemes between sequence lengths, with the model trained at 1 hour using 1 recording per batch and the model trained at 10.24 seconds using 352 different recordings. Hence, we also investigated an alternative scheme for comparison which is used in figures \ref{fig:wer_per_area_e22_epoch1}, \ref{fig:wer_per_epoch_per_seq} and \ref{fig:wer_per_area_epoch6} and table \ref{tab:floras50}. Here, the maximum sequence length is set to 1 hour for all sequence lengths, and the context size is controlled through by varying the local/windowed attention size. We found that both schemes resulted in similar results. All experiments are conducted on single 80GB A100s/H100s.

\section{Results and Analysis}
\begin{figure*}[tb!]
    \centering
    \includegraphics[width=18cm]{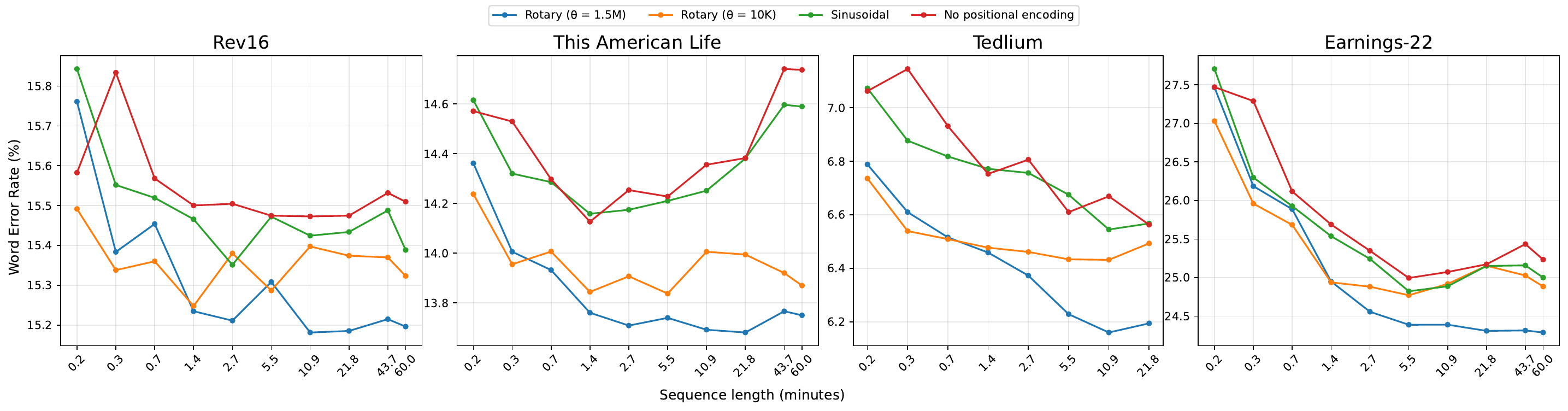}
    \caption{WER (\%) at each sequence length on each of the selected datasets for various positional encoding schemes.}
    \label{fig:posenccomparison}
\end{figure*}
\begin{figure*}[tb!]
    \centering
    \includegraphics[width=18cm]{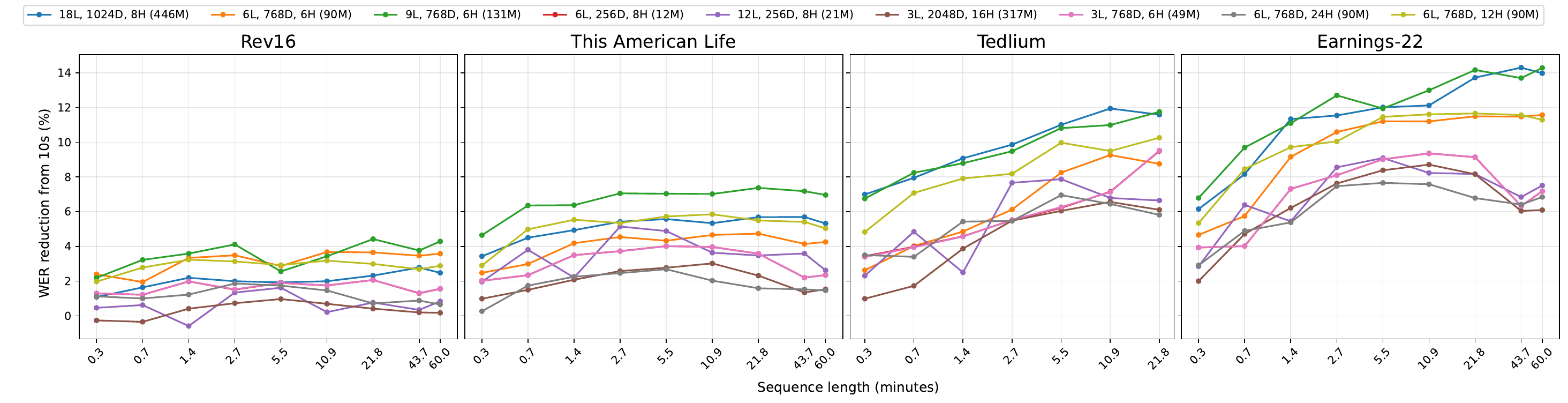}
    \caption{WER reduction (\%) from a baseline with 10 seconds of context, on each dataset. Positional encoding: Rotary ($\theta = 1.5$M).}
    \label{fig:pdecreaseacrossmodels}
\end{figure*}

The following section shows our experimental results. Each training run is repeated 3 times, and the average is reported. Unless otherwise specified, models are trained using rotary positional encodings \cite{su2021roformer} with $\theta =$ $1.5$M and a 6L, 768D, 6H model architecture (\textbf{L}ayers, hidden \textbf{D}imension, attention \textbf{H}eads).

\subsection{Evaluation Scheme}

\begin{table}[hbt]
\footnotesize
    \centering
    \begin{tabular}{ccccccc}
        \toprule
        \multirow{2}{*}{Evaluation Scheme} & \multicolumn{5}{c}{Stride (\% of sequence length)} \\
        & 100 & 75 & 50 & 25 & 12.5 & 6.25 \\ 
        \midrule
         Buffered & \textbf{27.6} & \textbf{26.2} & \textbf{26.2} & \textbf{26.2} & \textbf{26.2} & 26.2 \\
         Moving Averaged & \textbf{27.6} & 27.3 & 27.7 & 26.3 & \textbf{26.2} & \textbf{26.1} \\

        \bottomrule
    \end{tabular}
    \caption{Comparing WER (\%) when using different stride ratios for sequence length of 20s on Earnings-22 data}
    \label{tab:varyingstride}
\end{table}

When comparing the various different evaluations schemes discussed in $\S$ \ref{sec:contextfrag}, we find that they all lead to a comparable word error rate (WER) across all sequence lengths when using a stride equal to $12.5\%$ of the sequence length. This suggests that each method converges to a similar point where any context fragmentation is accounted for. Table \ref{tab:varyingstride} presents results when using varying strides for the buffered and moving averages schemes. SWA, which does not use a stride, results in a WER of 26.2\%. The results show that the buffered decoding converges to a similar performance to moving windowed decoding while using a much larger stride (and therefore less computation). However, for settings where the whole recording can fit into GPU memory, SWA is the most efficient method, therefore this setting will be used for all other evaluations in this paper. 

Experiments comparing the SWA scheme against a standard utterance-level evaluation that uses the utterance boundaries provided as part of Tedlium are also performed. The utterance level evaluation resulted in an WER of 7.1\%, and the windowed evaluation obtained an WER of 6.8\%. This result demonstrates that the context fragmentation caused by the utterance boundaries harms performance, with a 4.1\% relative WER increase compared to the windowed scheme. This effect would likely be even more impactful in real-world settings where utterance boundaries have to be made without human assistance.

\subsection{How Much Context Is Useful?}
\label{sec:howmuchcontext}

\begin{figure}[h!]
    \centering
    \includegraphics[width=6.5cm]{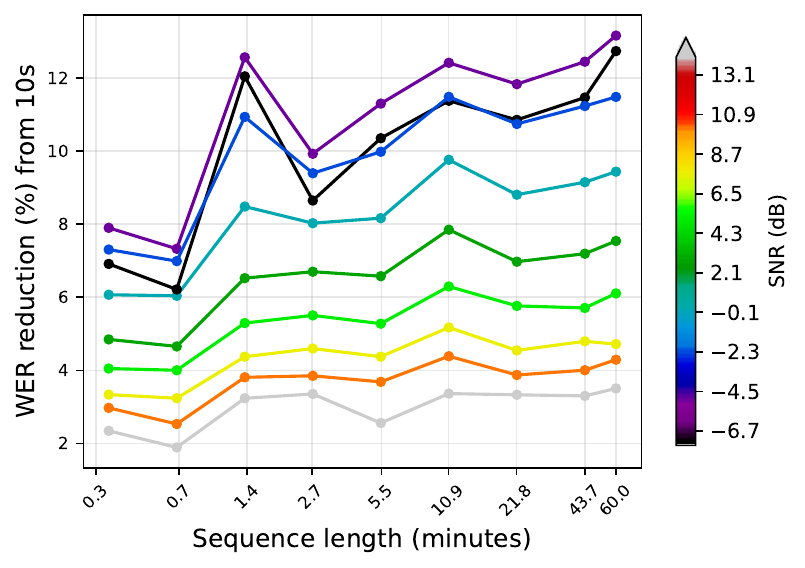}
    \caption{WER (\%) reduction from 10s baseline on Rev16 with varying levels of background music added as noise.}
    \label{fig:rev16noise}
\end{figure}

\begin{figure}[hbt]
    \centering
    \includegraphics[width=7cm]{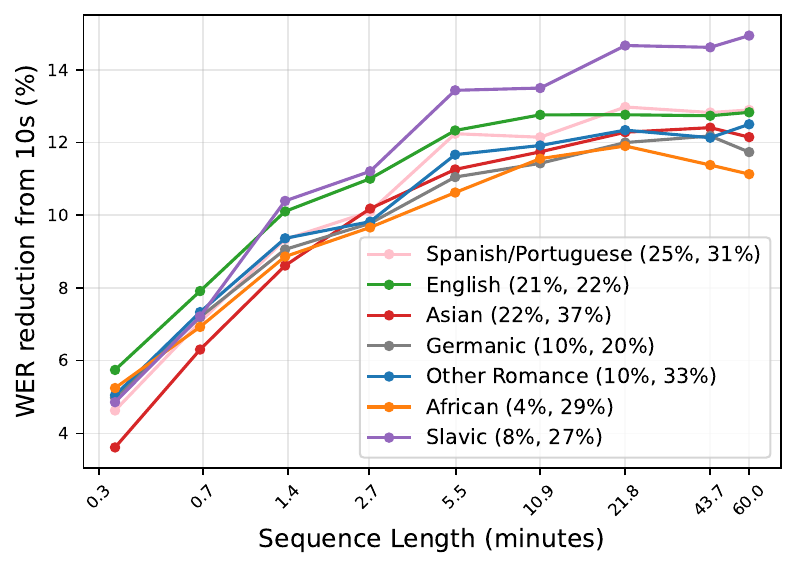}
    \caption{WER reduction (\%) from 10s of context grouped by L1 of speakers in Earnings22. (\% of total corpus, WER at 10s) is shown in the legend.}
    \label{fig:wer_per_area_e22_epoch1}

\end{figure}

Figure \ref{fig:posenccomparison} shows the WER at various context lengths on each of the datasets. On Earnings-22, our most challenging and out-of-domain dataset \ul{we find that up to 21.8 minutes of context is useful}, with a 11.5\% relative WER reduction from a baseline with 10s of context. Increasing the context beyond this leads to no further improvements. The majority of this improvement is reached with 5.5 minutes of context, with around an 11.2\% relative WER reduction compared to the baseline. 

\ul{We find that it is crucial to have at least 20 seconds of context}, with the largest improvement across all datasets happening between 10 and 20 seconds. Rev16 benefits the least from increasing the training/evaluation context, with little meaningful reduction in WER beyond 20–80 seconds of context. Rev16 is the most in-domain dataset (similar to the training data), consisting of podcasts from Apple Music. We find that the similarity to the training domain roughly corresponds to how much the models benefits from longer contexts. This can clearly be seen in figure \ref{fig:pdecreaseacrossmodels}. When evaluating on a subset of the training data, we see a similar pattern to Rev16, with no benefit from the longer context models. From manually inspecting the attention maps of the models, we observed that they attend to the full context that they are trained at, regardless of the dataset. Consequently, \ul{we believe that training with a longer context acts as an implicit bias, causing the model to learn solutions that use all of this context}. While these solutions are not necessarily more helpful in the training domain, these longer-context solutions may be more general, leading to better performance when evaluated out-of-domain. This is unexpected as we would expect the model to ignore context that is not helpful in the training domain.

As little benefit from longer-contexts was seen from in-domain data such as Rev16, we test whether artificially making Rev16 more out-of-domain through the addition of background noise, leads to an increased benefit from the longer context models. Results for this experiment, when noise is added at varying signal-to-noise ratios, are shown in figure \ref{fig:rev16noise}. The general trend shows that the models trained/evaluated with more context are more robust to higher levels of background noise. These results agree with our hypothesis that longer context models are learning solutions that generalize better to shifts in domain at test-time.

A breakdown of the WER for each of the L1 language families of speakers in Earnings-22 for each sequence length is provided in figure \ref{fig:wer_per_area_e22_epoch1} . Interestingly, we can see that the variance in performance between languages gets larger as the context size is increased. Certain language families such as Slavic consistently benefitted more from the additional context than the other language families. However, this is not necessarily a reflection of the accents, and is likely confounded by many other factors such as the conversation topic and training data distribution.

\subsection{Impact of Positional Encoding Method}
\label{sec:impactofpos}
From figure \ref{fig:posenccomparison} we can see the impact of varying the positional encoding scheme and the context length. The behaviour of each scheme as the context length is scaled differs depending on the evaluation dataset. For instance, when using the NoPos or the sinusoidal schemes, the WER starts to increase for context lengths greater than 1.4 minutes on the TAL dataset. This leads to the model performing worse with 1 hour of context than it does with 10 seconds. This may be due to This American Life featuring multiple acts within 1 hour, where the speakers and topic are changed. Additionally, TAL contains much more speakers overall than any other dataset. As NoPos or Sinusoidal schemes do not have an explicit distance bias built-in, the degradation here may be due to difficulty ignoring information from other speakers and topics.

Across all datasets, the rotary encoding method with $\theta = 1.5$M consistently demonstrates the best performance. This becomes more pronounced at longer context lengths, demonstrating that the positional encoding scheme is an important factor to consider when working with long contexts. In our investigations, different $\theta$ values did not yield significantly improved results from the values reported on at any sequence length.

\subsection{Impact of Model Size}
\label{sec:impactofmodelsize}

\begin{table}[h!]
    \centering
    \footnotesize
    \begin{tabular}{cccccc}
        \toprule
        Model Size & Rev16 & TAL & Tedlium & Earnings-22  \\\midrule
         6L, 256D, 8H (12M)  & 17.0 & 16.4 & 8.2 & 32.19 \\
         12L, 256D, 8H (21M) & 16.0 & 15.3 & 7.3 & 28.6 \\
         3L, 768D, 6H (49M)  &  17.0 & 14.6 & 7.0 & 28.3 \\
         6L, 768D, 6H (90M)  & 15.8 & 14.4 & 6.8 & 27.5 \\
         6L, 768D, 12H (90M) & 15.7 & 14.6 & 6.9 & 27.5 \\
         6L, 768D, 24H (90M) & 15.5 & 14.2 & 6.7 & 26.5 \\
         9L, 768D, 6H (131M)  & 15.4 & 14.2 & 6.7 & 26.4 \\
         3L, 2048D, 16H (317M) & 15.6 & 14.6 & 7.0 & 28.3 \\
         18L, 1024D, 8H (446M) & \textbf{14.8} & \textbf{13.5} & \textbf{5.9} & \textbf{24.8} & \\
         \bottomrule
    \end{tabular}
    \caption{Absolute WERs (\%) at 10s of context for each model size shown in figure \ref{fig:pdecreaseacrossmodels}.}
    \label{tab:absoluteweratmodelsize}
\end{table}

Given that many of the abilities seen in neural networks only arise at certain parameter scales \cite{wei2022emergent, brown2020language}, it is important to examine whether this is also the case for long context understanding. Therefore, models of varying widths, depths, and sequence lengths are also investigated, with the results presented in figure \ref{fig:pdecreaseacrossmodels}. Overall, the results show that models should be sufficiently deep and wide in order to consistently benefit from longer contexts. For example, the shallow but wide model, with 3L and 2048D, shows degradation at context lengths beyond 10 minutes on Earnings-22. Similar findings are seen for a deep and thin model with 12L and 256D. We find that having an attention head dimension of at least 64 is important for longer contexts. This can be seen from the 6H, 12H and 24H variants of the 6L 768D model, with the 24H model showing degradation when trained at longer sequence lengths. Smaller attention head dimensions likely lead to a less accurate similarity function, which may be more impactful for longer sequence lengths, when relevant information is sparser.

Table \ref{tab:absoluteweratmodelsize} demonstrates that continued scaling of model layers and depth does lead to an improved WER. However, there is no significant change in the relative improvement between a short context model and a longer context model as the model size is changed. These results align with the findings of \cite{olsson2022context} for language models, and can be seen by comparing the performance of the 9L 768D (131M) model with the 18L 1048D (446M). These models show a similar improvement from longer contexts despite around a 3.4$\times$ parameter count difference.

\subsection{Effect of Varying the Subsampling Factor}
\label{sec:ssfactor}
%

\begin{table}[hbt]
\footnotesize
    \centering
    \begin{tabular}{ccccccc}
        \toprule
        \multirow{2}{*}{Subsampling Factor} & \multicolumn{6}{c}{Sequence Length (minutes)} \\
        & 0.2 & 1.4 & 5.5 & 21.8 & 43.7 & 60.0 \\ 
        \midrule
         4$\times$ & 29.4 & 25.9 & 25.4 & 25.3 & 25.3 & OOM  \\
        8$\times$ & \textbf{27.5} & \textbf{25.0} & \textbf{24.4} & \textbf{24.3} & \textbf{24.3} & \textbf{24.3} \\

        \bottomrule
    \end{tabular}
    \caption{WER (\%) on Earnings-22 for different context sizes when varying the models subsampling factor.}
    \label{tab:sscomparison}
\end{table}

The primary experiments in this paper use a subsampling factor of 8$\times$. We also run experiments with a 4$\times$ subsampling factor, which are presented in table \ref{tab:sscomparison}. Other aspects of the subsampling setup are kept the same. As this results in higher memory usage, gradient checkpointing \cite{chen2016trainingsublinear} is used for all non-subsampling modules. Overall, the WER is roughly 3.5\% higher across all context lengths when using 4x subsampling. Most importantly, the model performance shows a similar trend for both the 4$\times$ and 8$\times$ subsampling across the different context lengths. \ul{Therefore, the plateau seen at higher context sizes is not due to the attention mechanism in/ability to process sequences beyond a certain length}. Ideally, we would like to also experiment with even larger subsampling factors e.g 16$\times$, however this is not straightforward with CTC based models as the number of output frames must always be greater than the number of tokens in the reference/target.

\subsection{Impact of Training Epochs}
\label{sec:impactofepoechs}
\begin{figure}[h!]
    \centering
    \includegraphics[width=8.9cm]{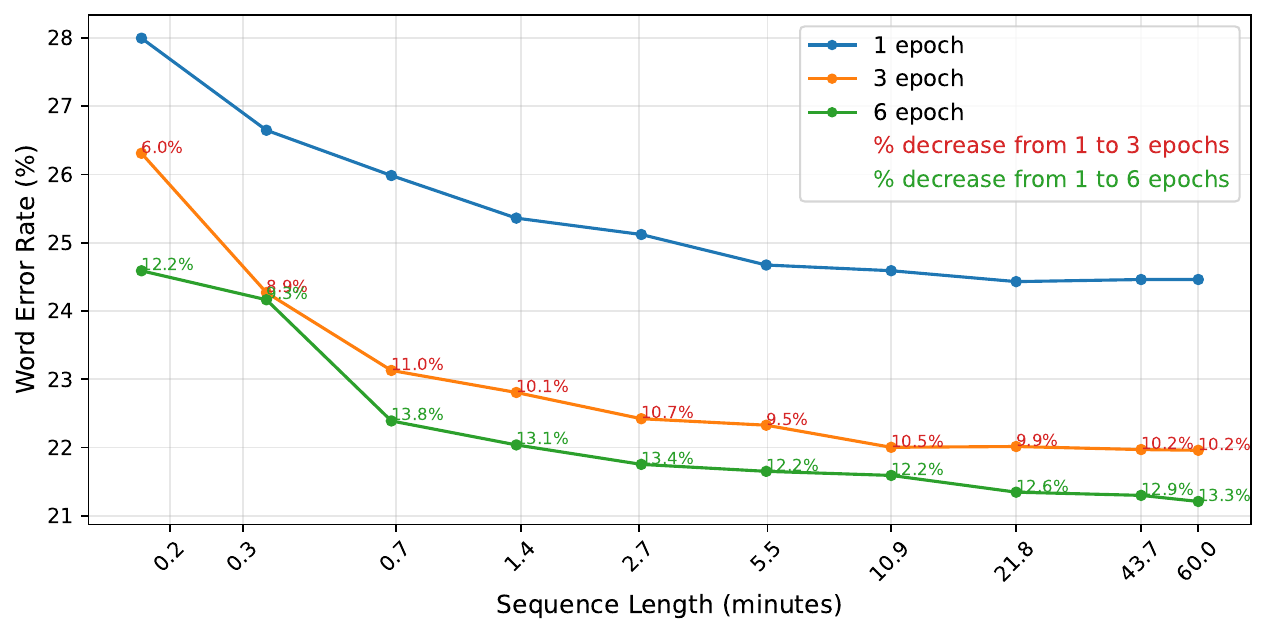}
    \caption{Effect of varying training epochs and attention window size on WER (\%). Results shown for Earnings-22 dataset.}
    \label{fig:wer_per_epoch_per_seq}
\end{figure}

\begin{figure}[hbt]
    \centering
    \includegraphics[width=7cm]{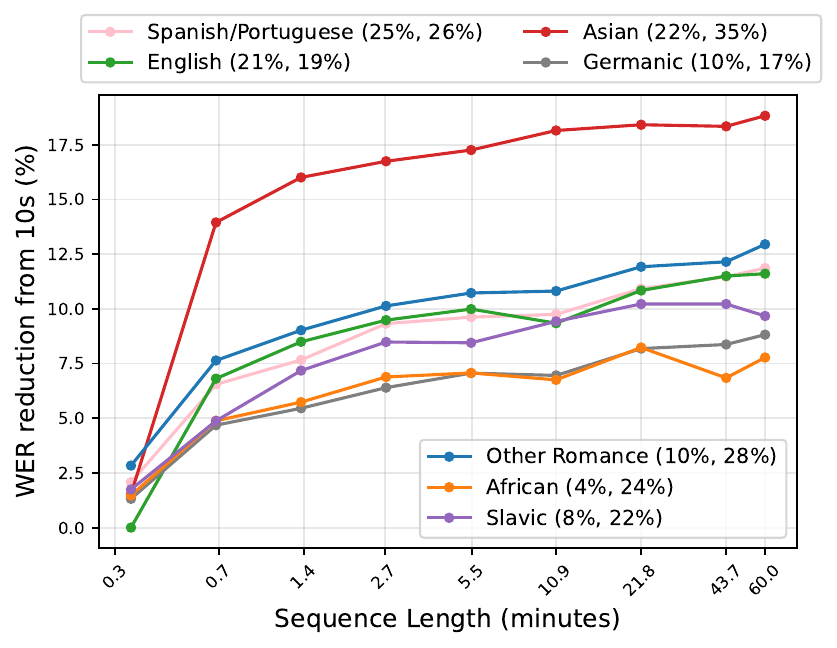}
    \caption{
    WER reduction grouped by L1 of speakers for Earnings22 dataset after 6 epochs. (\% of total corpus, WER at 10s).}
    \label{fig:wer_per_area_epoch6}
\end{figure}

The training time and number of repetitions/epochs may have an impact on the model's ability to use longer contexts. For example, \cite{singh2024transient} finds that in-context learning ability can degrade with training data repetition. To investigate if there exists a similar effect here, we conduct experiments where the total number of training epochs are varied between 1, 3, 6. 


The results in figure \ref{fig:wer_per_epoch_per_seq} shows that the models trained for different numbers of epochs show similar reductions in WER as the context is increased. This suggests that using 1 epoch for the majority of our experiments is sufficient when exploring the benefit of context. We can see that there would likely be little benefit from training for longer than 6 epochs (without introducing data augmentation techniques) from the smaller improvement between 3 and 6 epochs compared to 1 and 3. Interestingly, the results show that increasing the context size can be a more compute efficient way to reach a good performance than training for multiple epochs. For example, \ul{the model trained for 3 epochs at 10 seconds of contexts shows worse performance than the model trained for 1 epoch with 40 seconds of context}. Figure \ref{fig:conformerfastconformerspeed} shows that there is no significant increase in training time between these two sequence lengths. Similarly, the model trained for 6 epochs at 10 seconds shows a comparable performance to the model trained for 1 epoch with 5.5 minutes of context and worse performance than the 20 minutes of context model. 

The WER breakdown for native language families of speakers on Earnings-22 for the models trained for 6 epochs is shown in figure \ref{fig:wer_per_area_epoch6}. This differs considerably from the result for the 1 epoch setting in figure \ref{fig:wer_per_area_e22_epoch1}. Here the model shows a larger improvement from the context for L1 Asian speakers and for all other languages shows a considerably smaller improvement, compared to the 1 epoch setting. The model also finds the L1 Asian subset the most challenging in terms of absolute WER for both the 1 epoch and 6 epoch settings.

\subsection{Experiments on the Floras-50 Dataset}
\label{sec:floras50}

\begin{table}[hbt]

    \centering
    \begin{tabular}{cccccc}
        \toprule
        \multirow{2}{*}{Dataset} & \multicolumn{5}{c}{Sequence Length (minutes)} \\
        & 0.3 & 1.4 & 5.5 & 21.8 & 60.0 \\ \midrule
        Rev16 & 34.9 & 29.4& \textbf{28.9}& 31.2& 32.5\\
        This American Life & 15.6& 14.6& 14.0& \textbf{13.8}& 14.0\\
        Tedlium & 6.7& 6.7&6.7&\textbf{6.6}&N/A\\
        Earnings-22 & 24.1& 22.6& 22.3& \textbf{21.6}&21.8\\
        \bottomrule
    \end{tabular}
    \caption{WER (\%) on each dataset, for models trained at different context sizes, when using Floras-50 as training data.}
    \label{tab:floras50}
\end{table}

To investigate whether the observed performance trends generalize across training corpora, we conduct a set of experiments using the Floras-50 dataset. Models are trained on this corpus for six epochs at various sequence lengths and then evaluated. Floras-50 is approximately six times smaller than the Spotify dataset, so comparisons between the two should be interpreted cautiously. Results are shown in Table \ref{tab:floras50}. 

The results show that for all the datasets, there is a benefit from training and evaluating with at least 5.5 minutes of context. However, per-dataset performance trends differ considerably from those observed with the Spotify training corpus. Rev16 shows the largest improvement at 5.5 minutes, with a 17.7\% relative WER reduction compared to a 20-second baseline. Performance on this dataset degrades when the context exceeds 5.5 minutes. Furthermore, WER is high across all context lengths, at roughly twice the level observed for Spotify-trained models. This may be due to Rev16 consisting of podcast data that is highly similar to the Spotify corpus.

Performance on Tedlium shows little change, with only a 0.1\% absolute WER improvement when using 21.8 minutes of context. Interestingly, when models are evaluated after approximately three epochs (50\% of training), they exhibit different behaviour on Tedlium, with a larger 5.9\% relative WER reduction compared to a 20-second baseline. For the other datasets, performance trends at three epochs \textit{roughly} match those observed at six epochs in Table \ref{tab:floras50}. Both Earnings-22 and TAL show substantial improvements as context size increases, with best performance at 21.8 minutes of context. These improvements are larger than those observed for Spotify-trained models, reaching around a 10-11\% relative WER reduction at 21.8 minutes compared to a 20-second baseline.

Because all models exhibit some amount of degradation at one hour of context, we examined the distribution of recording lengths in the training data and found that recordings shorter than 10 minutes are far more common than those longer than 20 minutes. In contrast, the Spotify dataset exhibits a much more uniform length distribution. Based on this observation, we investigated whether an additional fine-tuning stage using only recordings longer than one hour would mitigate this degradation. However, this additional fine-tuning did not resolve the observed degradation.

Overall, results on this dataset further support the conclusion that context lengths of up to 21.8 minutes are beneficial during training and evaluation. We also observe larger gains from increased context when evaluating on podcast-based datasets (TAL, Rev16) using Floras-50 training data, compared to the Spotify training data. This helps supports our hypothesis that longer contexts are most beneficial under domain shift between training and evaluation. However, further investigation is needed to understand how changes in the training distribution affect behaviour at different context lengths.

\vspace{-0.5em}
\subsection{Impact of Altering the Context}
\label{sec:howusecontext}

To analyse the model's use of context, we alter the context beyond a buffer in various ways and measure the change in accuracy. Specifically, we take a 20-second window of the spectrogram and include an additional segment of the regular context of size $w_s$ on either side, resulting in a buffer size of $2 w_s + 20$. We then append randomly selected 20-second \textit{distractor} segments on either side, taken from a given source, until the audio has a length of 1 hour. Only the probabilities from the original 20-second window are used, and this window is moved until the entire recording has been transcribed. We use 4 different sources to sample the distractor segments from, which are as follows:
\begin{itemize}[label={}, leftmargin=*]
    \item \textbf{Within Dataset Context:} Segments are sampled from different recordings from the same dataset (Earnings-22)
    \item \textbf{Within Recording Context:} Segments are sampled from the same recording
    \item \textbf{Cross Dataset Context:} Segments are sampled from recordings in a different dataset (TAL)
    \item \textbf{No Context:} Segments are empty (contain zero values)
\end{itemize}
Earnings-22 is used as the dataset for this evaluation.  

\begin{figure}
    \centering
    \includegraphics[width=7cm]{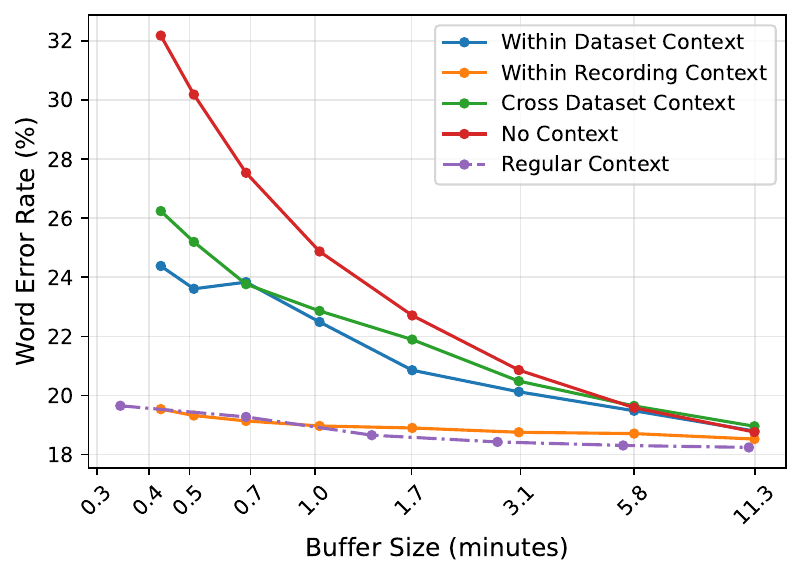}
    \caption{WER (\%) when context beyond the buffer size is replaced with random segments. }
    \label{fig:buffercontextexp}
\end{figure}

\begin{figure}
    \centering
    \includegraphics[width=7cm]{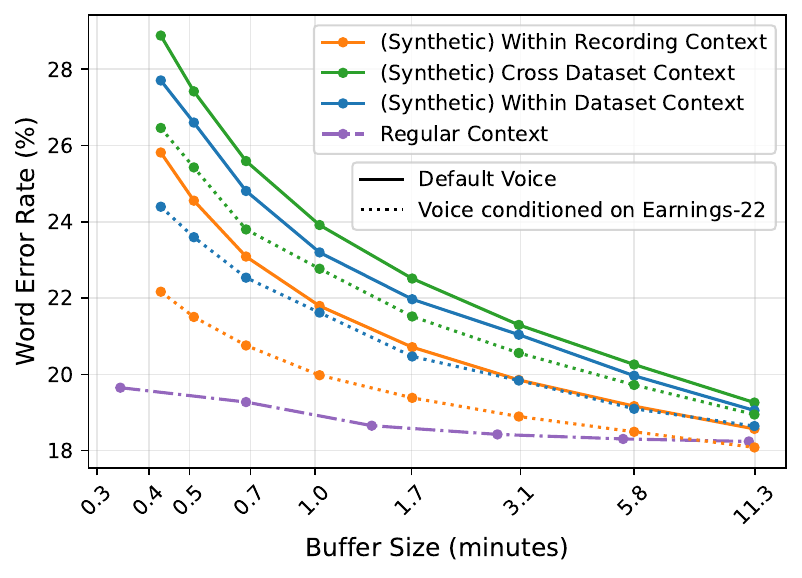}
    \caption{WER (\%) when context beyond the buffer size is replaced with \textbf{synthetic} random segments.}
    \label{fig:buffercontextexpsynthetic}
\end{figure}

The results as the buffer size is increased are shown in figure \ref{fig:buffercontextexp}. A \say{Regular Context} setting is also included as a baseline, which uses the models from figure \ref{fig:pdecreaseacrossmodels} which are trained/evaluate at a given context size without perturbing any of the context. From the difference in performance between the \say{Within Recording Context} and \say{Regular Context} settings, we can see that the ordering and content of the segments in the context does matter. For instance, at around 0.7 minutes these two settings show similar performance, despite the \say{Within Recording Context} having a full hour of unordered context available. However, the \say{Within Recording Context} setting is less harmful to performance than using context from other recordings from the same dataset. We hypothesise that this is due to the inclusion of different speakers, words, and phrases that are not relevant to current recording. The slight improvement, seen when using \say{Within Dataset Context} compared to the \say{Cross Dataset Context}, demonstrates that models are utilising some general features that are shared across different recordings from the same dataset. 

To gain further insight into the types of features in the context that the model is benefitting from, we create a synthetic version of each recording in Earnings-22 and TAL using a text-to-speech (TTS) system\footnote{{\url{https://huggingface.co/coqui/XTTS-v2}}} with the default voice. The experiments above are then repeated, using random segments from these synthetic datasets. As the voice used is the same across all recordings/datasets, this allows us to evaluate the purely linguistic component of the context. These experiments are presented in figure \ref{fig:buffercontextexpsynthetic}. The results show that the performance degrades the least when synthetic samples are taken from the same recording. This demonstrates that the model is using information about the linguistic distribution of the context to condition the outputs. As in figure \ref{fig:buffercontextexp} including context from the same dataset (but a different recording) is slightly less harmful than using context from a different dataset.

Figure \ref{fig:buffercontextexpsynthetic} also includes results where the synthetic recordings are generated by conditioning the samples from Earnings-22. All the settings show improved results when the TTS voice is more similar to the voice in the window, regardless of the linguistic content. This demonstrates that the model is using non-linguistic context to help adapt to the speaker's voice.


\subsection{Impact of Sudden Changes in the Recording}

\begin{figure}[hbt]
    \centering
    \includegraphics[width=7cm]{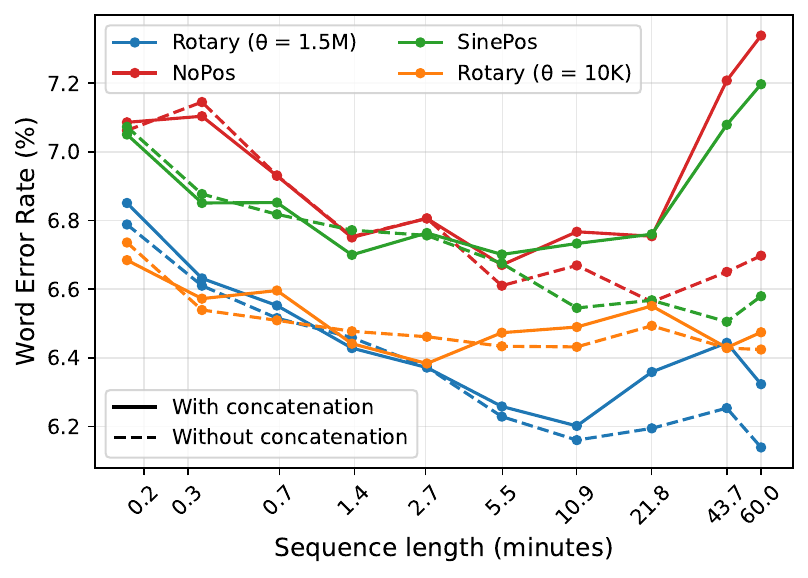}
    \caption{Comparing performance when other TED talks are included as context for the current recording (via concatenation).}
    \label{fig:concatTed}
\end{figure}

To simulate sudden changes in the content of the recording, we conduct additional experiments where recordings from TED-LIUM are randomly concatenated before and after the current recording. While this experiment is artificially constructed, similar context changes could exist in real-world recordings. For example, a recording of an oral session at a conference may contain different speakers talking back-to-back about their work. Therefore, it is important that long-context models are able to handle such situations.

From the results in figure \ref{fig:concatTed}, we can see that concatenating other TED talks causes an increase in WER across all the positional encoding methods. The NoPos and SinePos schemes are effected the most by concatenation, with the WER becoming worse at 60 minutes of context than at 10 seconds. We hypothesise that this is the same U shaped behaviour that is seen on the TAL dataset in figure \ref{fig:posenccomparison}. TAL features multiple acts within each show, where some speakers change and a different story is discussed. This demonstrates that the models have difficulty ignoring irrelevant information in the context. For the NoPos scheme, this behaviour is expected, as it provides no mechanism for distinguishing inputs that occur 5 minutes apart from those 20 minutes apart within the context. While rotary encodings show less degradation in the concatenated setting, they are still affected by this. Interestingly, we experimented with training a model where recordings are concatenated during the training process, to see if this improves the model's ability to deal with these sorts of context shifts. The results showed the same degradation to the original model, suggesting that this is due to some form of architectural shortcoming.


\vspace{-1em}
\subsection{In-Context Example Evaluation}
\label{sec:incontexteval}

\begin{figure}[hbt]
    \centering
    \includegraphics[width=7cm]{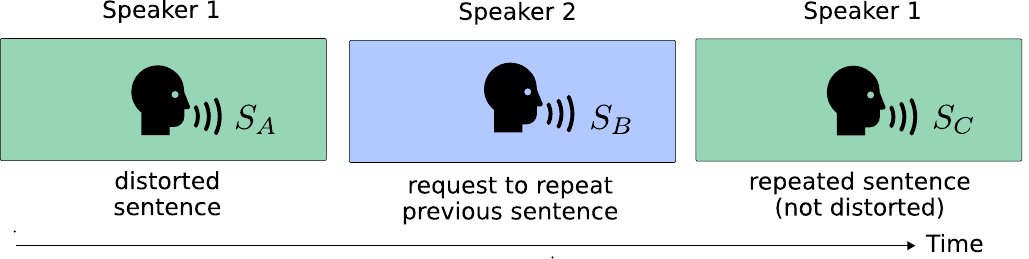}
    \caption{Depiction of the synthetic recording setup for the in-context evaluation}
    \label{fig:in-context}
\end{figure}

To test ASR model's ability to adapt based on contextual information, a small synthetic dataset is created. Recordings in this dataset feature two speakers $1$ and $2$ and use the following format (see figure \ref{fig:in-context}): An initial sentence $S_A$ is spoken by speaker 1, this sentence is corrupted or distorted or a subset of words are mispronounced. Then, speaker 2 requests speaker 1 to repeat the sentence (sentence $S_B$). Finally, speaker 1 repeats the initial sentence (sentence $S_C$), this time without distortion. Accuracy is measured by the model's ability to correctly transcribe a target word $T$ in $S_A$ given sentences $S_A$, $S_B$ and $S_C$ as context. Accuracy is also reported for when the repeated sentence $S_C$ is not provided as context. These settings are referred to as $A(T|S_A,S_B,S_C)$ and $A(T|S_A,S_B)$. 

Instances where the model is unable to transcribe $T$ correctly without the additional context of $S_C$, but correctly transcribes it when the repeated context $S_C$ is available, demonstrate the model is using this context to improve transcription of the first sentence. While this experiment does not evaluate the model's long-context ability, it examines how well the model can adapt based on explicit information provided in the context, which would also be useful at longer contexts. There are 20 recordings in total, each with a duration of around 20 seconds. We make the data publicly available here\footnote{\scriptsize https://github.com/robflynnyh/in-context-asr}. 

\begin{figure}[hbt]
    \centering
    \includegraphics[width=8.5cm]{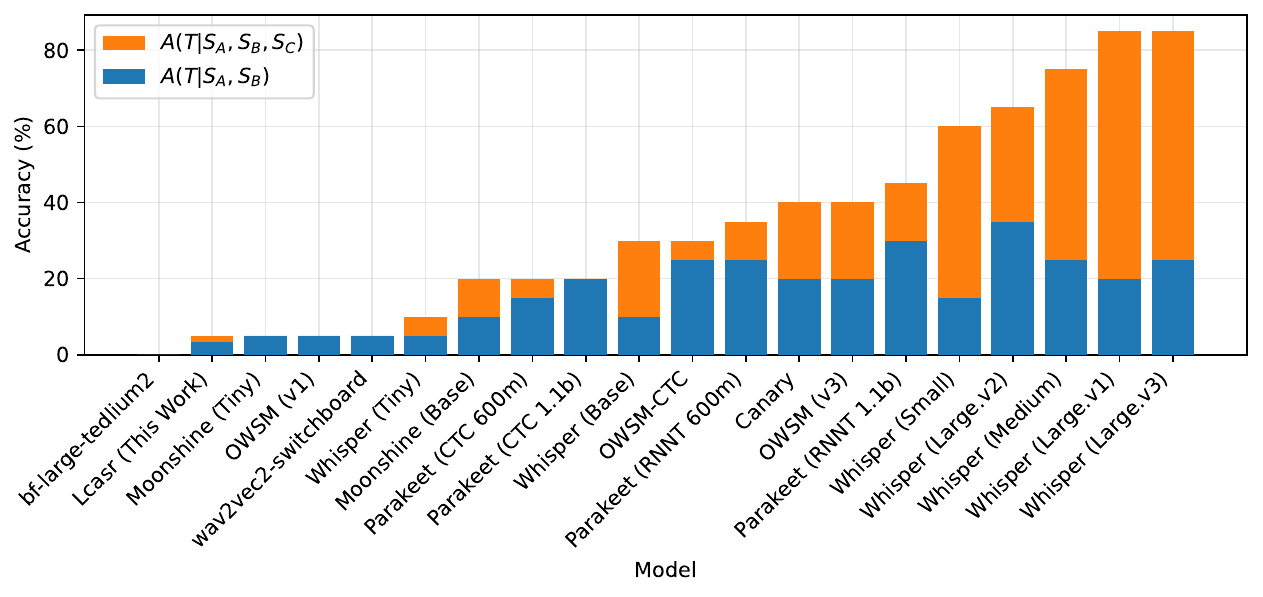}
    \caption{Results for the in-context example evaluation for a selection of openly available ASR models.}
    \label{fig:enter-label}
\end{figure}

For this dataset, we provide results for our own model (6L 768D 6H) from figure \ref{fig:pdecreaseacrossmodels}, as well as a range of openly available ASR models. The results demonstrate that none of the models are able to fully solve this task, despite correctly transcribing $S_B$ and $S_C$. This shows that the models either do not fully understand the meaning of $S_B$ and $S_C$ or have not learnt to adapt their predictions of $S_A$ based on this information. The Whisper models \cite{whisperradford2023robust}, all perform better than other equivalently sized models, and are the only models to perform better than 45\% accuracy, with the largest models achieving an accuracy of 85\%. For comparison, the similarly sized encoder-decoder models Canary\footnote{\scriptsize https://huggingface.co/nvidia/canary-1b} and OWSM v3 \cite{peng2023reproducing} models reach an accuracy of only 40\%. None of the models perform better than $35\%$ on the $A(T|S_A,S_B)$ setting, however this is expected, as many of the sentences are designed to be very difficult to transcribe without the repeated context. In this setting, Whisper large v2 performs better than v3. We \textit{hypothesise} that this may be due to the use of augmentation/regularization techniques that were used in v2 and not in v3 (based on the release notes).

In the $A(T|S_A,S_B,S_C)$ setting, CTC based model performed worse than equivalently sized transducer or encoder-decoder models. For example, the Parakeet CTC model performs $55.6\%$ worse than the RNNT version. Another trend that we notice is that performing more epochs on the same data tends to worsen model performance on this task. For example, Whisper Large v2 performs worse than v1, and v2 is trained for multiple epochs.

Our model showed fairly low performance on this task, however other large scale models such as OWSM v1 \cite{peng2023reproducing} and Moonshine Tiny \cite{jeffries2024moonshine} showed performance in the same range. Based on the models that perform well, we believe that data diversity, quality, and size may be particular important on this task. Our data is from one source (Spotify podcasts), so will show less diversity than a similarly sized corpus composed of heterogeneous sources. Additionally, pseudo labels are used in this work, which may impact performance. In future, we plan to extend this dataset and conduct more experiments to gain a better understanding into what aspects of training affect performance on this task.

\section{Conclusion}
In this work, we conducted a series of experiments to better understand the capabilities of current encoder-only attention-based ASR models in long-format settings. One key findings is that these models are generally able to benefit from up to 21.8 minutes of context—an order of magnitude larger than what is typically used. We found that selecting the appropriate positional encoding scheme, attention head dimension and number of layers were important at longer contexts. More context was particular beneficial when the shift between the training and testing domain was large, demonstrating that the additional context improves the robustness and generalisation of ASR models. By constructing a range of experiments using synthetic speech, we showed that the long-context ASR models are using both linguistic and acoustic features from the context. Additionally, we found that sudden changes in the context, such as a change in speaker, tend to worsen the performance of the models. In our future studies, we plan to investigate the long-context performance of other architecture types, such as encoder-decoders, as well as alternative operators to attention. We also plan to further investigate how the training data distribution and size affect long-context performance.


 
\bibliographystyle{IEEEtran}
\bibliography{mybib}

\end{document}